\documentclass{nsr}

\usepackage{amsmath,graphicx,array}
\usepackage{dcolumn,soul}

\usepackage{amsthm}
\usepackage[figuresright]{rotating}
\usepackage{algorithm, algorithmicx, algpseudocode}
\usepackage{listings}
\usepackage{hyperref}
\usepackage{bm}
\usepackage{braket}
\usepackage{cuted}

\jvol{XX}
\jnum{X}
\jyear{Year}
\doi{10.1093/nsr/XXXX}
\received{XX XX Year}
\revised{XX XX Year}
\accepted{XX XX Year}

\begin{document}

\dhead{RESEARCH ARTICLE}

\subhead{PHYSICS}

\title{Itinerant topological magnons and spin excitons in twisted transition metal dichalcogenides: Mapping electron topology to spin counterpart}

\author{Wei-Tao Zhou$^{1,\ddagger}$}

\author{Zhao-Yang Dong$^{3,\ddagger}$}

\author{Zhao-Long Gu$^{1,2,*}$}

\author{Jian-Xin Li$^{1,2,\dagger}$}

\affil{$^1$National Laboratory of Solid State Microstructures and Department of Physics, Nanjing University, Nanjing 210093, China}

\affil{$^2$Collaborative Innovation Center of Advanced Microstructures, Nanjing University, Nanjing 210093, China}

\affil{$^3$Department of Applied Physics, Nanjing University of Science and Technology, Nanjing 210094, China}

\authornote{\textbf{Corresponding author.} Email: waltergu@nju.edu.cn}
\authornote{\textbf{Corresponding author.} Email: jxli@nju.edu.cn}
\authornote{Equally contributed to this work}

\abstract[ABSTRACT]{Twisted transition metal dichalcogenide (tTMD) provides a highly tunable platform to explore the interplay between strong correlation and topology. Among them, the properties involving the charge degree of freedom have been extensively studied, while those related to spin are much less investigated. Motivated by the recent discovery of integer and fractional quantum anomalous Hall effects in tMoTe$_2$, for which the flat-band ferromagnetism is an essential prerequisite, we investigate theoretically the spin excitations out of the flat-band ferromagnetic ground state in tMoTe$_2$. Remarkably, we identify the itinerant magnons and spin excitons with nontrivial topology. We elaborate that the topology of these itinerant spin excitations, which are described as particle-hole bound states, inherits directly from that of the underlying electrons and is fundamentally different from that in local spin systems. Thus, we establish a direct relationship of the topology between the many-body excitations and their fundamental constituents. We further demonstrate that by tuning the displacement field, a topological transition for these excitations occurs, leading to a step-like change and bifurcation in the thermal Hall conductivity, which could serve as unique and compelling evidence to be tested experimentally. Our work deepens the understanding of spin excitations in itinerant electron systems with flat bands, provides a new paradigm to manipulate the magnon topology by electrical methods, and paves the way towards future investigation of magnonics in tTMD.}


\maketitle

\section{INTRODUCTION}\label{sec1}

tTMD moiré materials have emerged as a simple yet rich model system to study exotic phenomena, such as Mott insulators \cite{Regan2020Nature,Li2021Nature1,Regan2024NC}, generalized Wigner crystals \cite{Regan2020Nature, Xu2020Nature, Li2021Nature, Li2024Science}, unconventional superconductors \cite{Xia2025Nature, Guo2025Nature}, Kondo effect \cite{Zhao2023Nature, Guerci2023SciAdv} and many other correlated states of matter \cite{Wang2020NM, Tang2020Nature, Merkl2020NC, Huang2021NP, Gu2022NP, Anderson2023Science, Zhao2024NP, Kang2024Nature}. The intertwined correlation and nontrivial band topology have further induced the integer and fractional quantum anomalous Hall effects \cite{Li2021Nature2, Cai2023Nature, Park2023Nature, Xu2023PRX, Zeng2023Nature, Foutty2024Science}. In such cases, the tunable moiré flat bands stabilize itinerant ferromagnetism due to spontaneous spin polarization, which is a prerequisite for the emergence of the quantum anomalous Hall effects. Despite being universal and important in tTMD materials, the itinerant ferromagnetism, especially its spin excitations and the corresponding topology, have not been investigated.

Topological magnons are of significant importance due to their potential applications in non-dissipative magnonics. In the past, topological magnons were limited to local spin systems \cite{Zhang2013PRB, Mook2014PRB1, Mook2014PRB2, Owerre2016JPC, Laurell2018PRB, Seshadri2018PRB, Diaz2019PRL, Lu2021PRL, McClarty2022Annual, Santos2023NC}, where the magnon topology is induced by spin-spin exchange interactions, {\it e.g.}, the Dzyaloshinskii–Moriya interaction, and can be comprehensively understood within the framework of linear spin wave theory (LSWT). However, in such systems, the charge degree of freedom is frozen, making it difficult to manipulate the topological magnons using electrical means. On the contrary, itinerant magnons have the advantage that the spin is still coupled to the charge, offering the possibility of manipulating magnons by electrical means. But prior investigations on itinerant magnons have been limited to toy models \cite{Tasaki1992PRL, Tasaki1998PTP, DorettoPRB2015, Su2018PRB, Su2019PRB, Gu2019arXiv, Gu2021CPL} and their topological character has remained elusive due to the lack of an effective model for itinerant magnons.

In this work, we propose the realization of itinerant topological magnons in a realistic tMoTe$_{2}$ system, whose topology can be tuned by applying a displacement field. By describing the itinerant magnons as particle-hole bound states, we propose and prove that the topology of the itinerant magnons is induced by a direct electron hopping, thus inherits directly from that of the underlying electrons. This mechanism is fundamentally distinct from that in local spin systems. Moreover, we find itinerant topological spin excitons in a range of twist angles as the result of the cooperation of the Hubbard interaction and the gap between the electron bands, and demonstrate that their topology also stems from the underlying electrons and can be tuned by the displacement field. We suggest that the topological transitions of both the itinerant magnons and spin excitons can be experimentally probed by the spectral measurement and that of magnons can be probed by a step-like rise and bifurcation of the thermal Hall conductivity.

\begin{figure*}
    \centering
    \includegraphics[width=\linewidth]{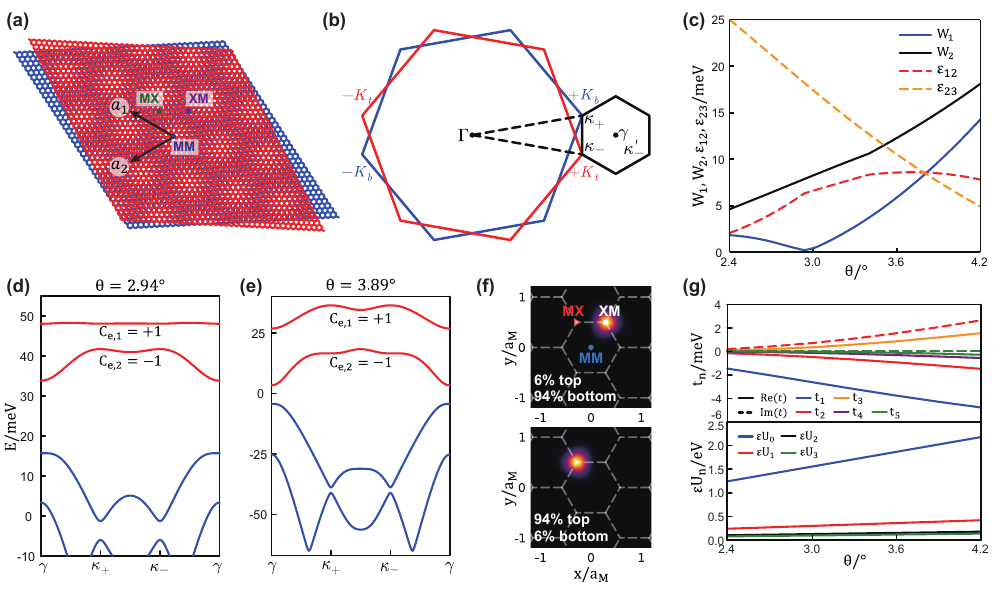}
    \caption{
    \textbf{Moiré superlattice and electronic tight-binding model.} (\textbf{a}) Moiré superlattice. $a_{1}, a_{2}$ are moiré superlattice vectors. MM, MX, XM are high symmetry stacking regions. (\textbf{b}) Brillouin zones of the top (red) and bottom (blue) monolayer and the moiré Brillouin zone (black). (\textbf{c}) Widths of the topmost two bands $W_{1}$, $W_{2}$ and the first two band gaps $\varepsilon_{12}$, $\varepsilon_{23}$. (\textbf{d})(\textbf{e}) Moiré band structure at $\theta=\text{2.94}^\circ$ and $\theta=\text{3.89}^\circ$, respectively. The topmost two bands in both cases have nonzero Chern numbers $\pm 1$. (\textbf{f}) Two generated Wannier states at $\theta=\text{2.94}^\circ$, which center at XM, MX stacking regions and show large layer polarization. (\textbf{g}) Calculated hopping parameters and gate-screened interaction strengths.
    }
    \label{fig:Model}
\end{figure*}

\begin{figure}
    \centering
    \includegraphics[width=\linewidth]{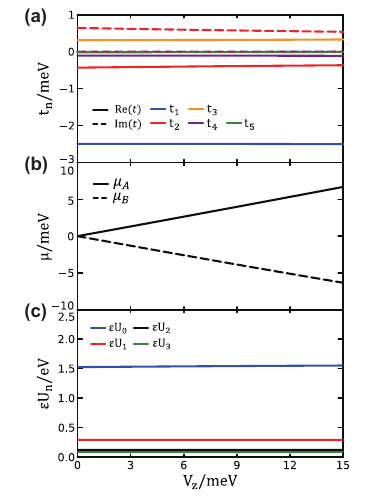}
    \caption{
    \textcolor{red}{\textbf{Dependence of tight-binding parameters on displacement field $V_{z}$}. (\textbf{a}) The hopping parameters, (\textbf{b}) the staggered onsite chemical potentials at sublattice A and B and (\textbf{c}) the  interaction strengths with the displacement field from $V_{z}=0$ to $V_{z}=15$ meV at $\theta=2.94^\circ$, respectively.}
    }
    \label{fig:DependenceOnVz}
\end{figure}

\section{ELECTRONIC MODEL}\label{sec2}

We start with the continuum model of tTMD, in which the low-energy band structure is described by the topmost valence bands of the two monolayers at $\pm K_{b,t}$ points [see Fig.~\ref{fig:Model}\textbf{a} and \textbf{b}], modulated by the moiré potential and interlayer tunneling \cite{Wu2019PRL, Pan2020PRR, Devakul2021NC}. Due to the large spin-orbit coupling, these topmost valence bands exhibit quadratic dispersion resembling that of a free electron with a modified mass and are spin-valley locked. Since the $\pm K$ valleys are decoupled due to large momentum transfers while related by the time-reversal symmetry, we focus on the $+K$ valley (spin-$\uparrow$ block), of which the moiré Hamiltonian is given by
\begin{equation}
    \mathcal{H}_\uparrow = \begin{pmatrix}
        -\frac{\hbar^2 (\bm{k}-\bm{\kappa}_+)^2}{2m^*}+\Delta_+(\bm{r}) & \Delta_T(\bm{r}) \\
        \Delta^\dagger_T(\bm{r}) & -\frac{\hbar^2 (\bm{k}-\bm{\kappa}_-)^2}{2m^*}+\Delta_-(\bm{r})
    \end{pmatrix}.
    \label{eq1}
\end{equation}
This matrix is defined in the layer bases. $m^*$ is the effective mass of an electron. The layer-dependent momentum offset $\kappa_{\pm}$ captures the rotation of the two monolayers in momentum space, as shown in Fig.~\ref{fig:Model}\textbf{b}. $\Delta_{\pm}$, $\Delta_T$ are the moiré potential and interlayer tunneling, respectively. Constrained by the space group $D_3$ and time-reversal symmetry $\mathcal{T}$ of tTMD, $\Delta_{\pm}$, $\Delta_T$ can be parameterized by
\begin{equation}
    \Delta_\pm(\bm{r}) = 2V\sum_{j=1,3,5} \cos(\bm{g}_j \cdot \bm{r} \pm \psi),
    \label{eq2}
\end{equation}
\begin{equation}
    \Delta_T(\bm{r}) = w(1+e^{i\bm{g}_2 \cdot \bm{r}}+e^{i\bm{g}_3 \cdot \bm{r}}),
    \label{eq3}
\end{equation}
where $\bm{g}_1=(4\pi/(\sqrt{3}a_M),0)$ ($a_M$ is the moiré lattice constant) and $\bm{g}_j$ with $j=2, 3, ..., 6$ are related to $\bm{g}_1$ by $(j-1)\pi/3$ rotation. Both $\Delta_{\pm}$ and $\Delta_T$ are truncated to the lowest harmonic order, which is sufficient to capture the low-energy electron bands. When applying the displacement field $V_{z}$, $\Delta_{\pm}$ is replaced by $\Delta_{\pm} \pm V_{z}/2$. $(V,\psi,w)$ specify the strength and distribution of moire potential as well as the strength of the interlayer tunneling. Here we take m$^*$ to be 0.6$m_0$ ($m_0$ is the mass of a free electron) and $(V,\psi,w)$ to be ($20.8$ meV, $107.7^\circ$, $-23.8$ meV), which are from a large-scale first-principle calculation of AA-stacked tMoTe$_2$ \cite{Wang2024PRL}.

In Fig.~\ref{fig:Model}\textbf{c}, we present the electronic band widths and gaps derived by the continuum model. It shows that the first band reaches extreme flatness at $2.94^\circ$, which serves as the magic angle in tTMD \cite{Devakul2021NC, Morales2024PRL} and resembles that in twisted bilayer graphene \cite{Bistritzer2011PNAS, Santos2012PRB}. In the following, we focus on the $2.94^\circ$ case and choose a commensurate twist angle $3.89^\circ$ as a comparison. In Fig.~\ref{fig:Model}\textbf{d} and \textbf{e}, we show the electron band structures in the two cases, of which the first two bands have nonzero Chern numbers $\pm 1$ and are separated from the rest of the spectrum. We generate an effective tight-binding model by constructing two layer-polarized Wannier states which emerges as an effective honeycomb lattice as shown in Fig.~\ref{fig:Model}\textbf{f}, and the calculated hopping parameters and interactions are shown in Fig.~\ref{fig:Model}\textbf{g}. The effective electronic model can be well captured by the generalized Kane-Mele model \cite{Kane2005PRL1, Kane2005PRL2} with the Hubbard interaction:
\begin{equation}
    \label{eq4}
    \begin{split}
        H_{\text{TB}} &= t_1 \sum_{\braket{ij}\sigma} c^{\dagger}_{i\sigma}c_{j\sigma} + |t_2| \sum_{\braket{\braket{ij}}\sigma} e^{i\nu_{ij} \sigma \phi}  c^{\dagger}_{i\sigma}c_{j\sigma} \\ & + U \sum_{i} n_{i\uparrow} n_{i\downarrow} + T_{ij}.
    \end{split}
\end{equation}
Here, $\braket{ij}$ and $\braket{\braket{ij}}$ denote the nearest neighbor (NN) and next NN, respectively, $\nu_{ij}= +1(-1)$ if the electron makes a left (right) turn to get to the NN site, $\sigma=\pm 1$ for spin-$\uparrow$ (spin-$\downarrow$) and $\phi=2\pi/3$. 
$T_{ij}$ denotes the hopping terms beyond $t_2$ that have small but non-negligible amplitudes (truncated up to $t_5$).

\textcolor{red}{As shown in Fig.~\ref{fig:DependenceOnVz}, we present the dependence of the tight-binding parameters on displacement field $V_{z}$. It shows that $V_{z}$ mainly introduces staggered chemical potentials on different sublattices while slightly changes the hopping parameters and interaction strengths. This consequence is reasonable considering that the two Wannier states have large layer polarization which would feel opposite electrical potentials and in the small $V_{z}$ region, their distribution is robust against $V_{z}$. It further proves that the effective Kane-Mele-Hubbard model is a good start point to investigate the low-energy physics of tTMD at small twist angles.}

One can consult Supplementary or Ref.~\cite{Devakul2021NC} for more details of the continuum model, the generation of Wannier states and the computation of the hopping and interaction parameters.

\begin{figure*}
    \centering
    \includegraphics[width=\linewidth]{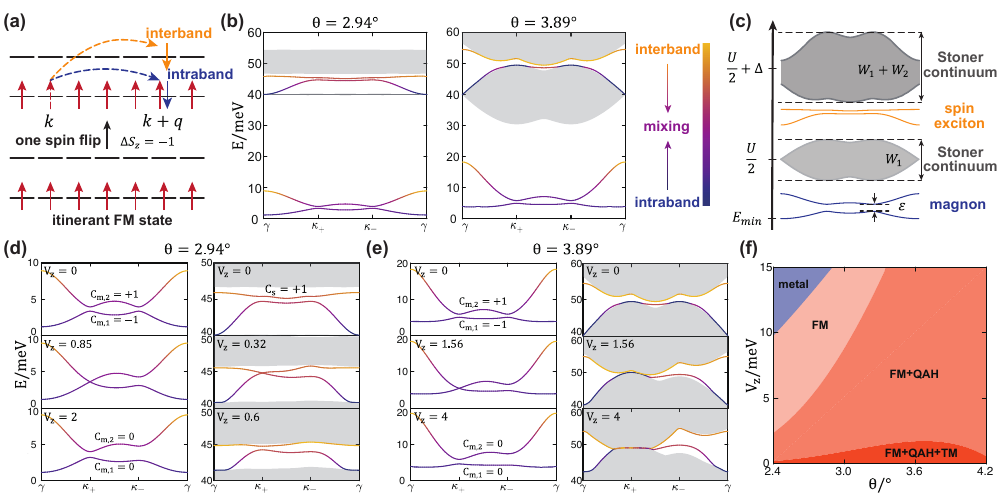}
    \caption{
    \textbf{Spin excitations out of the itinerant FM state.} (\textbf{a}) Schematic of the itinerant FM ground state, and the intraband and interband spin-flip particle-hole excitations. (\textbf{b}) Spectra of spin excitations at $\theta=\text{2.94}^\circ$ (left) and $\theta=\text{3.89}^\circ$ (right) without a displacement field. The color of magnons and spin excitons represents the component of the intraband and interband spin flips. (\textbf{c}) Schematic of the typical energy bands for spin excitations, which is divided into magnons, spin excitons and Stoner continua. $U$, $\Delta$, $W_{1}$, $W_{2}$, $\varepsilon$ denote the Hubbard interaction, average gap between the topmost two electron bands, widths of the two electron bands and the gap between the magnon bands, respectively. (\textbf{d})(\textbf{e}) Evolution of magnons (left) and spin excitons (right) at $\theta=\text{2.94}^\circ$ and $\theta=\text{3.89}^\circ$ by varying $V_{z}$. (\textbf{f}) Phase diagram with respect to the twist angle and displacement field $V_{z}$. All results in (\textbf{b})(\textbf{d})(\textbf{e})(\textbf{f}) are obtained at $U=\text{80 meV}$.
    }
    \label{fig:Magnon}
\end{figure*}

\section{ITINERANT tOPOLOGICAL MAGNONS AND SPIN EXCITONS}\label{sec3}

\textcolor{red}{In tTMD systems, the itinerant ferromagnetic (FM) ground state has been experimentally detected by a magnetic hysteresis loop at partial filling in Ref.~\cite{Anderson2023Science,Cai2023Nature,Park2023Nature,Xu2023PRX}. From a theoretical perspective, the emergence of this fully spin-polarized state can be attributed to the fact that the strong-correlation effect overwhelms that of kinetics in flat-band systems, as the kinetic-energy gain by equally occupying both spin up and down states from low to high energies cannot compensate the interaction penalty by the overlap of electrons with different spins\cite{Tasaki1992PRL, Tasaki1998PTP, DorettoPRB2015, Su2018PRB, Su2019PRB, Gu2019arXiv, Gu2021CPL}.}

Now, we investigate the spin excitations out of the itinerant FM ground state at filling $\nu=+1$, {\it i.e.}, one hole per moiré unit cell. 
Generally, the spin excitations in an itinerant system come from the electronic spin-flip particle-hole excitations. For convenience we perform a particle-hole transformation and refer a hole (an electron) as an electron (a hole). The FM ground state is expressed by $\ket{\text{GS}}=\prod_{\bm{k}\in 1\text{BZ}} \alpha^\dagger_{\bm{k}l\uparrow} \ket{0}$, where $\alpha^\dagger_{\bm{k}l\uparrow}$ creates a spin-$\uparrow$ electron on the lower band with momentum $\bm{k}$. The Hilbert space with one spin flip ($\Delta S_{z} = -1$) is spanned by $\ket{\bm{k},\bm{q},\mu}=\alpha^\dagger_{\bm{k}+\bm{q}\mu \downarrow} \alpha_{\bm{k}l\uparrow} \ket{\text{GS}}$, where $\mu=l(u)$ denotes the lower (upper) band. This space contains both intraband and interband spin flips, as shown in Fig.~\ref{fig:Magnon}\textbf{a}. Due to the translation symmetry, the total momentum $\bm{q}$ is a good quantum number. By acting $H_{\text{TB}}$ on the $\Delta S_{z}=-1$ space we obtain the Hamiltonian matrix $H(\bm{q})$ at each $\bm{q}$ with its element $H_{\bm{k}\mu,\bm{k}'\mu'}(\bm{q})=\bra{\bm{k},\bm{q},\mu} H_{\text{TB}} \ket{\bm{k}',\bm{q},\mu'}$. In the presence of inversion symmetry, we express $H(\bm{q})$ to be the following compact form
\par
\begin{strip}
    \begin{equation}
        \centering
        H(\bm{q}) = \sum_{\bm{k}\mu=l,u} \left[ \frac{U}{2} + \varepsilon_{\mu}^\downarrow(\bm{k}+\bm{q}) - \varepsilon_{l}^\uparrow(\bm{k}) \right] \ket{\bm{k},\bm{q},\mu} \bra{\bm{k},\bm{q},\mu} - \frac{U}{N} \sum_{n=A,B} \ket{\bm{q},n} \bra{\bm{q},n},
    \label{eq5}
    \end{equation}
\end{strip}
where $\varepsilon^{\sigma}_{\mu}$ with $\mu=l(u)$ is the dispersion of the lower (upper) spin-$\sigma$ electron band, $n=A, B$ denote the two sublattices of the effective honeycomb lattice, $N$ is the number of unit cells, and $\ket{\bm{q}, n} \equiv \sum_{\bm{k}}c^{\dagger}_{\bm{k}+\bm{q}n\downarrow}c_{\bm{k}n\uparrow}\ket{\text{GS}}$. Here, $c^{\dagger}_{\bm{k}n\sigma}$ creates a spin-$\sigma$ electron of sublattice $n$ with momentum $\bm{k}$. When the inversion symmetry is explicitly broken by a displacement field, which equals to introducing an opposite chemical potential at different sublattices, $H(\bm{q})$ has a slightly more complex form, which is shown in Supplementary.

The spin excitations with momentum $\bm{q}$ are calculated by the exact diagonalization (ED) of $H(\bm{q})$, which are performed on a $N=72\times72$ k-mesh (such a large-scale ED calculation is possible as the $\Delta S_{z}=-1$ space scales linearly in $N$). The spectra of spin excitations along the high-symmetric direction in the moiré Brillouin zone are shown in Fig.~\ref{fig:Magnon}\textbf{b} for $\text{2.94}^\circ$ and $\text{3.89}^\circ$, which are obtained with a Hubbard interaction $U=80$ meV being experimentally accessible by tuning the relative dielectric constant $\varepsilon$ to be about $20$.

Overall, the excitation spectra consist of four parts: two continua, two well-defined excitations at low energies below $10$ meV ($\text{2.94}^\circ$) or 20 meV ($\text{3.89}^\circ$), and two well-defined excitations in the intermediate energies between the two continua. The differences in spectral features between the two twist angles are: 1) for a small twist angle $\text{2.94}^\circ$, the lower continuum has an almost vanishing width, 2) for a large twist angle $\text{3.89}^\circ$, both continua expand in width so that the excitations between them merge into the continua in most $\bm{q}$ ranges where the merged excitations lose the well-defined feature.

To understand the nature of these spin excitations, we resort to a qualitative analysis by inspecting each term in Eq.~\ref{eq5}. The first term describes the scattering of the state $\ket{\bm{k},\bm{q},\mu}$ which involves the particle-hole excitation of a spin-$\downarrow$ electron and a spin-$\uparrow$ hole with the momentum $\bm{k}+\bm{q}$ and $\bm{k}$, respectively. For a fixed $\bm{q}$, the excitation band is broadened by the internal momentum $\bm{k}$ into a continuum, which is exactly the Stoner continuum---a universal character of the itinerant system. Thus, from the first term of Eq.~\ref{eq5}, we determine that the lower continuum which consists of intraband flips, locates at $U/2$ with a width of $W_{1}$ and the higher continuum which consists of interband flips, locates at $U/2+\Delta$ with a width of $W_{1}+W_{2}$, where $\Delta$, $W_{1}$, $W_{2}$ is the average gap and widths of the two electron bands, respectively. The second term of Eq.~\ref{eq5}, on the contrary, describes the scattering of the bound states formed by particle-hole pairs. By a Fourier transformation $\ket{\bm{q},n}=\sum_{i} e^{i\bm{q} \cdot \bm{R}_{i}} c^{\dagger}_{in\downarrow} c_{in\uparrow} \ket{\text{GS}}$, we see that $\ket{\bm{q},n}$ is a collection of the onsite spin flips $\beta^{\dagger}_{in} \equiv c^{\dagger}_{in\downarrow} c_{in\uparrow}$ projected onto the itinerant FM ground state, which is in essence the magnons and spin excitons.

All the qualitative features based on Eq.~\ref{eq5} are summarized in Fig.~\ref{fig:Magnon}\textbf{c}. A comparison between Fig.~\ref{fig:Magnon}\textbf{c} and the ED results in Fig.~\ref{fig:Magnon}\textbf{b} confirms that the two well-defined excitations at low energies are two magnons corresponding to the acoustic and optical branches arising from the two-sublattice structure of the honeycomb lattice, two well-defined excitations at intermediate energies are two spin excitons and the others are Stoner continua. In this way, we can also understand that the lower continuum with a vanishing width at $\text{2.94}^\circ$ is due to the extreme flatness of the electron band as shown in Fig.~\ref{fig:Model}\textbf{d}. We note that in contrast to the usual charge excitons, the identified spin excitons are lifted in energy by $U/2$ arising from the itinerant FM background.

Now, we turn to the discussion of the topological properties of the itinerant magnons and spin excitons. When there is no displacement field $V_{z}=0$, we find that the Chern numbers of the two gapped magnon bands are $C_{m}=\pm 1$, as shown in Fig.~\ref{fig:Magnon}\textbf{d} and \textbf{e}. Therefore, the itinerant magnons have nontrivial topology. Interestingly, we find that the topology of magnons can be tuned by the displacement field $V_{z}$. Increasing $V_{z}$, the gap between two magnon bands decreases, and eventually closes at the $\kappa_+$ point when $V_{z}\approx 0.85$ meV for $\text{2.94}^\circ$ and $V_{z}\approx 1.56$ meV for $\text{3.89}^\circ$. With a further increase of $V_{z}$, the closed gap reopens, but now the Chern numbers of both magnon bands become zero. As for the spin excitons, we note that the high-quality flatness of the electron band, {\it i.e.}, the small ratio of $W_{1,2}/\Delta$, plays an essential role for the appearance of the well-defined spin excitons, as shown in Fig.~\ref{fig:Magnon}\textbf{b}. So, at the magic angle $\theta=\text{2.94}^\circ$ where the well-defined spin exciton exists, we can find that the upper spin exciton has a nonzero Chern number $C_{s}=+1$ for $V_{z}=0$, and a transition to the topological trivial state accompanying with a gap close-and-reopen occurs as $V_z$ increases (see Fig.~\ref{fig:Magnon}\textbf{d}). In the cases away from the magic angle such as $\text{3.89}^\circ$, the broadening of the topmost flat electronic bands (Fig.~\ref{fig:Model}\textbf{e}) leads to a large $W_{1,2}/\Delta$, so that the spin excitons would merge into the Stoner continua and there is no possibility to consider their band topology.

In Fig.~\ref{fig:Magnon}\textbf{f}, we present a comprehensive $\theta$-$V_{z}$ phase diagram \textcolor{red}{derived both from the analysis of electron and magnon bands}. The boundary of the metal and general FM phase is determined by \textcolor{red}{the instability of magnons (the minimum energy of magnons reaching zero)}, which is closely related to the width of the first electron band $W_{1}$ (see Extended Data Fig.~2). The general FM phase can be further divided into a conventional FM state (FM) with trivial electron and magnon topology, an FM state with nontrivial electron topology but trivial magnon topology resulting in a quantum anomalous Hall effect (FM+QAH), and an FM state with both nontrivial electron and magnon topology (FM+QAH+TM). From the phase diagram, one can clearly see that the displacement field as an electrical means can control the magnon topology. It is expected that this phase diagram will stimulate and facilitate the experimental explorations of the magnetic excitations discussed in this paper.

\begin{figure*}
    \centering
    \includegraphics[width=\linewidth]{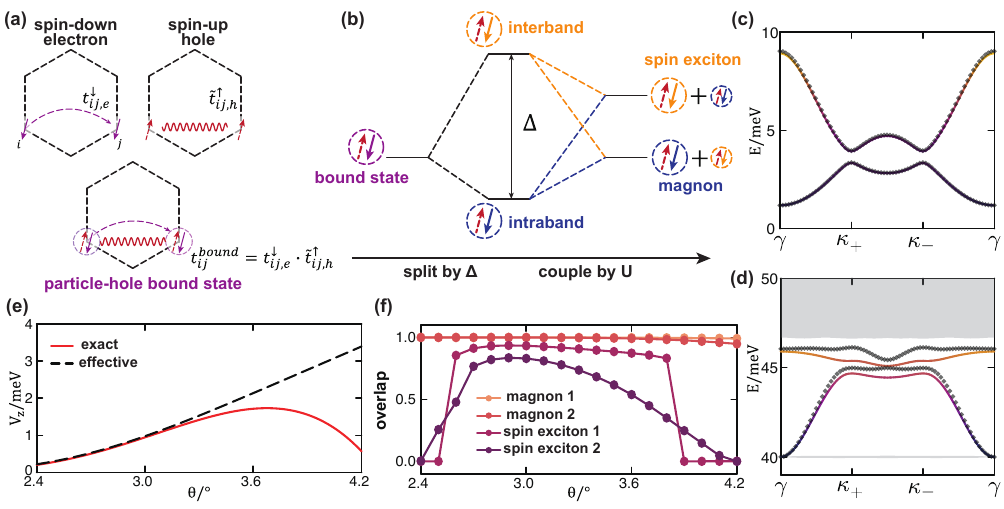}
    \caption{
    \textbf{Origin of magnon and spin exciton topology.} (\textbf{a}) Illustration of the direct inheritance of magnon topology from electronic topology. The magnon is described as a particle-hole bounded pair in an itinerant magnetic systems, of which the hopping is the product of those of a spin-$\downarrow$ electron and a spin-$\uparrow$ hole, with the latter renormalized by the itinerant FM background. (\textbf{b}) Magnetic effective model. The particle-hole bounded pair is split into its intraband and interband projections by the gap between the two electron bands $\Delta$. The two parts are further coupled by the Hubbard interaction, so that one evolves into two magnons and the other into two spin excitons. (\textbf{c})(\textbf{d}) Comparison of the magnon dispersion in (\textbf{c}) and spin exciton dispersion in (\textbf{d}) obtained by exact diagonalization (colored lines) and magnetic effective model (diamond lines) at $(\theta, V_{z}, U)=(\text{2.94}^\circ, \text{0}, \text{80 meV})$. (\textbf{e}) Comparison of magnon topological transition boundary. (\textbf{f}) Overlap of the states obtained by the exact diagonalization and magnetic effective model at $\kappa_+$ point.
    }
    \label{fig:TopoOrigin}
\end{figure*}

\section{MAGNETIC EFFECTIVE MODEL AND ORIGIN OF TOPOLOGY}\label{sec4}

In local spin systems, the magnonic excitations are well understood as bosons in the framework of LSWT, and their nontrivial band topology is usually induced by the Dzyaloshinskii–Moriya (DM) interaction acting as the vector potential for the propagation of free magnons. However, in the case of itinerant magnets, the lack of an exact local spin per physical site, along with the absence of the corresponding effective model describing the spin wave excitations, leaves the understanding of itinerant topological spin excitations rather vague. 

As shown in Fig.~\ref{fig:Magnon}\textbf{c}, the two continua and the spin excitons scale with $U$, while the magnons are located at rather low energies. So we first consider the simplest case focusing only on the magnons, which is equivalent to the large $U$ limit, pushing the continua and the spin excitons to infinite energies. Given that double occupancy is excluded in this limit, we introduce a magnetic effective model consisting of the projected onsite spin flip operators $\tilde{\beta}^{\dagger}_{in} \equiv \beta^{\dagger}_{in}\ket{\text{GS}}\bra{\text{GS}}$, based on an analogy with the LSWT where the local spin flip is mapped into a hard-core boson,
\begin{equation}
    H_{\text{mag}} = \sum_{ij} t_{ij}^{\text{bound}} \tilde{\beta}^{\dagger}_{i} \tilde{\beta}_{j}
    \label{eq6}
\end{equation}
where the sublattice index is absorbed into the site index, $t_{ij}^{\text{bound}} \equiv \bra{\text{GS}}\beta_{i} H_{\text{TB}} \beta^{\dagger}_j \ket{\text{GS}}$ is the hopping of the projected onsite spin flip, which corresponds to the hopping of a magnon in the itinerant FM systems. In fact, the extraction of the magnetic effective model equals to reducing the $\Delta S_{z}=-1$ space to $B_2 \equiv \{ \ket{\bm{q},A},\ket{\bm{q},B} \}$ and the Hamiltonian $H(\bm{q})$ to a 2$\times$2 matrix with its element $H_{mn}(\bm{q})=\bra{\bm{q},m} H_{\text{TB}} \ket{\bm{q},n}$, with $m, n = A, B$. By diagonalizing the reduced Hamiltonian matrix, we perfectly reproduce the magnon dispersion (see Extended Data Fig.~3), which confirms the fact that the nature of the itinerant magnon is the onsite spin flip $\beta^{\dagger}_{in}$ projected onto the itinerant FM ground state. As sketched in Fig.~\ref{fig:TopoOrigin}\textbf{a}, $t_{ij}^{\text{bound}}$ can be directly expressed as the product of the hopping of a spin-$\downarrow$ electron $t^{\downarrow}_{ij,e}$ and an effective hopping of a spin-$\uparrow$ hole $\tilde{t}^{\uparrow}_{ij,h}$ (proof can be found in Supplementary), so the nonzero phase factor of the next NN hopping of electrons would enter into that of magnons. The magnetic effective model thus resembles a bosonic version of the Haldane model \cite{Haldane1988PRL} which has staggered flux (complex $t_{2}^{\text{bound}}$) while the net flux is zero (real $t_{1}^{\text{bound}}$). It demonstrates unambiguously that the topology of itinerant magnons inherits directly from the topology of electrons.

We point out that the origin of the magnon topology in itinerant systems is essentially different from that in local spin systems. The former is induced by a direct electron hopping depending on $t$ but irrelevant to $U$, while the latter comes from the effective spin-spin interaction, {\it e.g.}, the DM interaction, which is a second-order virtual process depending on $t^2/U$. We also note that the effective $\tilde{t}^{\uparrow}_{ij,h}$ stems from the charge fluctuations of the FM ground state (see Supplementary), which is only nonzero in itinerant systems but vanishes in local spin systems. With this magnetic effective model, we can also understand the tuning effect of displacement field $V_{z}$ on the topology of magnons. A nonzero $V_{z}$ breaks the inversion symmetry and introduces an effective staggered onsite energy for magnons in Eq.~\ref{eq6}, which can change the topology of magnons similar to the famous Haldane model \cite{Haldane1988PRL}.

In the realistic circumstance in this paper where $U$ is finite (equal to $80$ meV) and the electronic energy gap $\Delta$ not too small (about 10 meV), we can include the spin excitons in our magnetic effective model by considering the effect of $\Delta$. As depicted in Fig.~\ref{fig:TopoOrigin}\textbf{b}, $\Delta$ splits the bound states (purple circle) into the intraband (blue circle) and interband (yellow circle) projections. Concurrently, a finite $U$ couples these two parts, resulting in the formation of low-energy magnons (primarily the intraband part with a minor interband contribution) and high-energy spin excitons (primarily the interband part with a minor intraband contribution), both of which are bound states. This partitioning of magnons and spin excitons aligns with the exact results presented in Fig.~\ref{fig:Magnon}. In order to account for all these physical processes, it is necessary to construct a magnetic effective model based on the enlarged bases $B_{4} \equiv \{ P_{l}\ket{\bm{q},A},P_{u}\ket{\bm{q},A},P_{l}\ket{\bm{q},B},P_{u}\ket{\bm{q},B} \}$, which are composed of the split bound states $P_{l,u} \beta^{\dagger}_{in} \ket{\text{GS}}$. Here, $P_{l}$ ($P_{u}$) is the projection operator onto the lower (upper) electron band. In this way, one can quantitatively check the effectiveness of the magnetic effective model and physical picture sketched in Fig.~\ref{fig:TopoOrigin}\textbf{b}. By diagonalizing the 4$\times$4 Hamiltonian matrix constructed in the $B_{4}$ space, we can obtain the eigenvalues $E_{\bm{q}}$ and eigenstates $\phi_{i}(\bm{q})$. In Fig.~\ref{fig:TopoOrigin}\textbf{c}-\textbf{d} and Fig.~\ref{fig:TopoOrigin}\textbf{e}, we present the obtained results for the spin excitation spectra $E_{\bm{q}}$ and the boundary of the topological transition of magnons, respectively, compared with those calculated from ED. In Fig.~\ref{fig:TopoOrigin}\textbf{f}, the overlap $P_{n}(\bm{q}) \equiv \sum_{i}|\langle\psi_{n}(\bm{q})|\phi_{i}(\bm{q})\rangle|^2$ of the eigenstates are shown, where $\psi_{n}(\bm{q})$ are those obtained by the ED calculation. An overall comparison shows clearly that in the range of the twist angles $\text{2.7}^\circ\sim\text{3.3}^\circ$, the description of the magnons based on the magnetic effective model coincides perfectly with the ED calculation, and the overlap of the spin excitons between the two methods is over 0.75. It confirms that the physical picture for the magnon and spin exciton based on the magnetic effective model (Fig.~\ref{fig:TopoOrigin}\textbf{b}) is reasonable. Based on this picture, one can see that the spin excitons evolve from the projected onsite spin-flip states, which acquire a nonzero phase factor in their hoppings inheriting from that in electrons. Therefore, the topology of spin excitons also stems from the topology of electrons.

\begin{figure*}
    \centering
    \includegraphics[width=\linewidth]{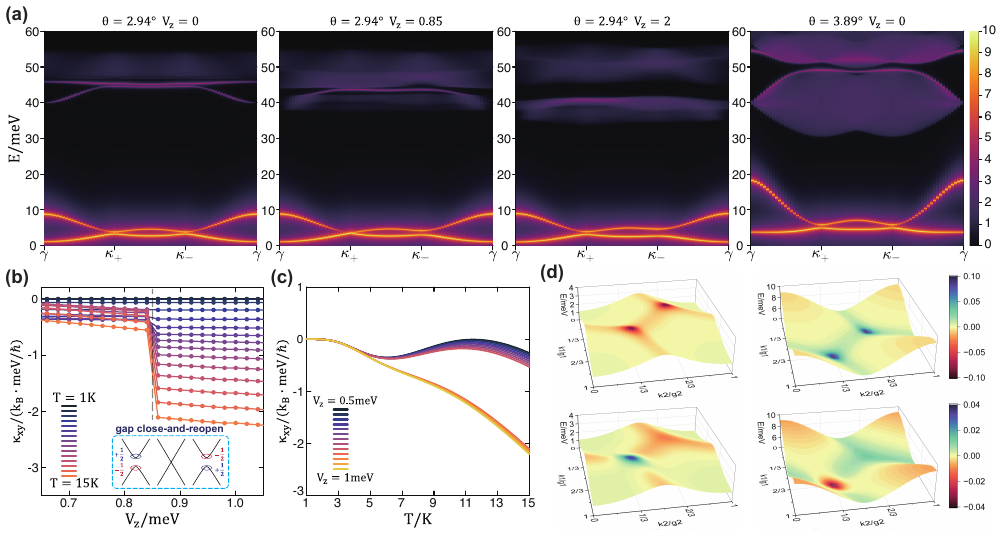}
    \caption{
    \textbf{Dynamical spin structure factor and thermal Hall conductivity.} (\textbf{a}) Logarithmic heatmap of the dynamical spin structure factor. From left to right, the first three panels are calculated at $\theta=\text{2.94}^\circ$, $V_{z}=\text{(0, 0.85, 2) meV}$, $U=\text{80 meV}$ and the forth panel at $\theta=\text{3.89}^\circ$, $V_{z}=\text{0}$, $U=\text{80 meV}$, respectively. (\textbf{b}) Thermal Hall conductivity of magnons as a function of $V_{z}$. A step-like rise of thermal Hall conductivity emerges at the magnon topological transition. Inset sketches the Berry curvature distribution at $\kappa_{+}$ right before and after the topological transition. (\textbf{c}) Thermal Hall conductivity of magnons as a function of $T$. A bifurcation appears at moderate temperatures as a signal of magnon topological transition. (\textbf{d}) Berry curvature distribution of the two magnon bands in momentum space at $\theta=\text{2.94}^\circ$, $V_{z}=\text{0}$ (top two panels) and $\theta=\text{2.94}^\circ$, $V_{z}=\text{2 meV}$ (bottom two panels).
    }
    \label{fig:ThermalHall}
\end{figure*}

\section{DYNAMICAL SPIN STRUCTURE FACTOR}\label{sec5}

We have established that topological magnons and spin excitons exist in tTMD with an itinerant FM ground state in a wide range of twist angles and a transition to trivial excitations can be tuned by adjusting the displacement field. Considering that the tuning of the displacement field is the popular and unique way in tTMD to adjust continuously the electronic structure, the tTMD provides a competitive system to probe the topological properties of these spin excitations. 

The first experimental observable is the dynamical spin structure factor, which reflects both the dispersion and the spectral weight of the excitations. Via the definition of the bosonic spin operator $\beta^{\dagger}_{\bm{q}n}$, we can calculate the spin structure factor $-\frac{1}{\pi} \text{Im}G^{R}(\bm{k},\omega)$ with the retarded Green's function $G^{R}(\bm{k},\omega)$ given by,
\begin{equation}
    G^{R}(\bm{q},\omega) = \sum_{mn} \frac{|\bra{\psi_{m}(\bm{q})} \beta^{\dagger}_{\bm{q}n} \ket{\rm GS}|^2}{\omega-\omega_{m\bm{q}}+i\delta}
    \label{eq7}
\end{equation}
where $m$ sums over all the eigenstates $\ket{\psi_{m}(\bm{q})}$ with total momentum $\bm{q}$ in $\Delta S_{z}=-1$ space and $\omega_{m\bm{q}}$ is the corresponding eigen energy, $n=A,B$ the sublattice index, $\beta^{\dagger}_{\bm{q}n}$ the Fourier transformation of $\beta^{\dagger}_{in}$. The results for two twist angles $\text{2.94}^\circ$ and $\text{3.89}^\circ$ are shown in Fig.~\ref{fig:ThermalHall}\textbf{a}. Overall, most spectral weights concentrate on the magnons and spin excitons, especially the two magnon bands. So, it facilitates the probe of these bound states. For the magic angle $\text{2.94}^\circ$ and $V_{z}=0$, where the topmost electron band becomes extremely flat, the lower Stoner continuum almost does not show up and the upper one also has a weak spectra weight. In the case of $\text{3.89}^\circ$, where the topmost electron band becomes much more dispersive, the two Stoner continua acquire more weight. In this case, the obvious bright edges of Stoner continua in fact reflect the weight of the spin excitons, but it is difficult to distinguish them from the continua. So, the test of the topological magnon and spin exciton experimentally is more suitable around the magic angle $\text{2.94}^\circ$. We note that, with the increase of $V_{z}$, a linear Dirac dispersion of the magnon bands emerges at $\kappa_+$ point when the gap between two magnons vanishes, {\it i.e.}, at the magnon topological transition, which can be probed by a spectral weight detection.

\section{THERMAL HALL CONDUCTIVITY}\label{sec6}

In recent years, the thermal Hall conductivity has emerged as a practical and insightful physical quantity for detecting topological states and characterizing the topological phase transitions in magnetic systems. In this regard, we study the thermal Hall conductivity of the topological spin excitations and its variation with the displacement field. As discussed above, the spin exciton is in fact fragile to the change of displacement field and it also situates in a high-energy region which has negligible effect on the thermal Hall conductivity at low temperatures, so we focus only on the contribution of magnons. The thermal Hall conductivity $\kappa^m_{xy}$ can be calculated using linear response theory \cite{Matsumoto2014PRB} and it has the following form
\begin{equation}
    \label{eq8}
    \kappa^m_{xy} = -\frac{k_B^2T}{\hbar}\sum_{n\bm{k}}\left( c_2\left[ g(\varepsilon_{n\bm{k}}) \right] -\frac{\pi^2}{3} \right)\Omega_{n\bm{k}}.
\end{equation}
where the band index $n$ sums over the two magnon bands, $g(\varepsilon)$ is the Bose-Einstein distribution function and $c_2(x)=\int^{x}_{0}dt\left( \text{ln}\frac{1+t}{t} \right)^2$, and $\Omega_{n\bm{k}}$ is the Berry curvature of the $n$th magnon band at momentum $\bm{k}$. In Fig.~\ref{fig:ThermalHall}\textbf{b}, we show $\kappa^m_{xy}$ as a function of $V_{z}$ for different temperature $T$. Firstly, we need to point out that there is no quantized behavior of $\kappa^m_{xy}$ as that of electrons, due to the bosonic character of spin excitations. Starting from the lowest temperature $T=1$ K, one can see that there is essentially no thermal Hall conductivity in the whole $V_{z}$, though the magnons are topologically nontrivial for $V_{z}$ below 0.85 meV. With the gradual increase of $T$, a noticeable finite $\kappa^m_{xy}$ appears and grows with $V_{z}$, but it always exhibits flat $\kappa^m_{xy}$ with $V_{z}$ until $T=5$ K. When $T>5$ K, a step-like rise in the magnitude of $\kappa^m_{xy}$ occurs at the topological transition point where the gap between two magnons closes and reopens. The same phenomenon manifests itself in the temperature dependence of $\kappa^m_{xy}$ as shown in Fig.~\ref{fig:ThermalHall}\textbf{c}. The thermal Hall conductivity shows stark distinct $T$-dependence before and after topological transition via the tuning of the displacement field $V_{z}$ and exhibits an obvious bifurcation. We propose that these two remarkable dependences of $\kappa^m_{xy}$ on $V_{z}$ and $T$, as compelling evidence on the topological transition we proposed here, can be used as an experimental test. 

To explore the physical reason of these phenomena, we note that the essential factor determining the behavior of the thermal Hall conductivity is the distribution of the Berry curvature in momentum space, which plays a similar role to the density of states in the usual electron conductivity. Hence, we show the calculated results of $\Omega_{n\bm{k}}$ in Fig.~\ref{fig:ThermalHall}\textbf{d} for the two magnon bands at $\theta=\text{2.94}^\circ$ with $V_{z}=\text{0}$ and $V_{z}=\text{2 meV}$, respectively. It shows that the magnitude of $\Omega_{n\bm{k}}$ increases gradually from $\gamma$ point to $\kappa_+$ and $\kappa_-$ points, and forms two peaks around $\kappa_+$ and $\kappa_-$. For $V_{z}=\text{0}$, $\Omega_{n\bm{k}}$ has the same sign for the acoustic and optical magnons and the same magnitude at $\kappa_+$ and $\kappa_-$ points. On the other hand, for $V_{z}=\text{2}$ meV where the two reopened magnon bands have zero Chern numbers, each of the two magnons exhibits opposite $\Omega_{n\bm{k}}$ in momentum space. This explains why the magnon bands have nonzero Chern number at $V_{z}=\text{0}$ and zero Chern number at $V_{z}=\text{2}$ meV, as the total contribution of the Berry curvature to the Chern number cancels out in the latter case. Having the knowledge of the Berry curvature, we are now at the position to understand the phenomena shown in Fig.~\ref{fig:ThermalHall}\textbf{b} and \textbf{c}. At low temperatures, only the low-energy magnons around $\gamma$ can be thermally excited, they have tiny contribution to the thermal Hall conductivity. With the increase of temperature, those magnons near the $\kappa_+$ and $\kappa_-$ points carrying the majority of Berry curvature are involved gradually. The step-like rise in $\kappa^m_{xy}$ across the topological transition can be attributed to the quantized change of the Berry curvature from $\mp 1/2$ to $\pm 1/2$ at $\kappa_+$ \cite{Berry1984}, where the gap closes and reopens, as shown in the inset of Fig.~\ref{fig:ThermalHall}\textbf{b}. Weighted by the prefactor $c_{2}[g(\epsilon)]$, the magnitude of the step-like rise depends on temperature. In the case of $\text{3.89}^\circ$ (see Extended Data Fig.~4), accompanying the similar step-like change across the transition, a sign reversal of thermal Hall conductivity also emerges, which is due to the detail distribution of the Berry curvature in momentum space. Because of the same physical reason, the temperature dependence of thermal Hall conductivity also exhibits noticeable bifurcation when the displacement field $V_z$ crosses the transition point.

Finally, we note that as an itinerant FM system, the total thermal Hall conductivity includes both contributions of magnon and electron bands. However, it is expected that the step-like and bifurcation features unveiled above would not be blurred by the electronic contributions $\kappa^e_{xy}$. First, the phase diagram Fig.~\ref{fig:Magnon}\textbf{f} shows that the magnon topological transition occurs at a far smaller $V_z$ than that of electron (about 7 meV at $\text{2.94}^\circ$). The Berry curvature of electron bands changes little at that small $V_{z}$, so does $\kappa^e_{xy}$. Second, as $\kappa^e_{xy}$ is proportional to the electron Hall conductivity $\sigma^e_{xy}$ to a temperature dependent factor, one can filter out $\kappa^e_{xy}$ from the total $\kappa_{xy}$ by detecting $\sigma^e_{xy}$.

\section{SUMMARY AND DISCUSSION}\label{sec7}

In summary, we theoretically predict the existence of itinerant topological magnons and spin excitons in a large range of twist angles in tMoTe$_2$. We demonstrate that the topology of the itinerant magnons and spin excitons can be tuned by displacement field $V_z$, and the topological transition can be experimentally identified by a step-like rise in the $V_z$ dependence and a bifurcation in the temperature dependence of thermal Hall conductivity. We elaborate that the topology of itinerant magnons and spin excitons originates directly from that of electrons, in that the hopping of magnons and spin excitons as particle-hole bound states, acquires a flux from that of the underlying electrons. This mechanism differs from that in local spin systems, where the topological magnons are typically induced by the Dzyaloshinskii–Moriya interaction.

Our work has raised the following prospects. First, it is worth emphasizing that itinerant topological magnons and spin excitons are ubiquitous in the topological flat-band systems with itinerant ferromagnetic ground states. tTMD currently provides an ideal platform to explore these exotic excitations due to its high tunability in terms of topology, bandwidth and electron filling. The occurrence of these phenomena is expected to extend beyond tTMD, as they can potentially be observed in any material system with the topological flat-band ferromagnetic ground state.
Second, the magnetic effective model describing itinerant magnons as particle-hole bound states, can serve as a good starting point to theoretically investigate the spin excitations in itinerant FM systems, which may not be restricted to integer filling per unit cell. In the flat-band limit, we can easily inspect the evolution of itinerant magnons by continuously varying the filling factor, assuming an appropriate itinerant FM ground state.
Finally, we propose that the exotic spin excitations and topological transition could be probed by spectral measurements and heat transport experiments, which are still rare in the investigations of tTMD. We believe that the theoretical predictions presented here can inspire further advancements in such experiments in tTMD.
Our work deepens the understanding of spin excitations in itinerant electron systems with flat bands, provides a new paradigm to manipulate the magnon topology by electrical methods, and paves the way for future experiments investigating magnonics in tTMD.

{\it Note added.}-- After completing this work, we learned that a related study by W.-X. Qiu {\it et al.} \cite{Qiu2025arXiv} also find theoretically the topological magnons in tMoTe$_2$.

\section*{DATA AVAILABILITY}
The data that support the findings of this study are available from the corresponding authors upon request.

\section*{SUPPLEMENTARY DATA}
Supplementary data are available at NSR online.

\section*{ACKNOWLEDGMENTS}
We thank J.-S. Wen, L. Wang and L.-W. He for helpful discussions. 

\section*{FUNDING}
The work was supported by National Key Projects for Research and Development of China (Grant No. 2021YFA1400400, 2024YFA1408104), the National Natural Science Foundation of China (Grant No. 12434005, No. 92165205 and No. 12404174), and Natural Science Foundation of Jiangsu province (Grant No. BK20230765).
We thank e-Science Center of Collaborative Innovation Center of Advanced Microstructures for support in allocation of CPU.

\section*{AUTHOR CONTRIBUTIONS}
Z.-L. Gu and J.-X. Li conceived the project. W.-T. Zhou carried out the numerical calculations with the help of Z.-Y. Dong and Z.-L. Gu. W.-T. Zhou, Z.-Y. Dong, Z.-L. Gu and J.-X. Li performed the theoretical analyses and wrote the manuscript. J.-X. Li supervised the work.
\\\\
\noindent{\it \textbf{Conflict of interest statement.} None declared.}

\bibliographystyle{nsr}
\bibliography{mainref}

\end{document}